\documentclass[preprint,english,showpacs,11pt,floatfix]{revtex4}

\usepackage{babel,amsmath,amssymb,dcolumn}

\usepackage[dvips]{graphics}

\usepackage{hyperref}

\usepackage{caption2}

\usepackage{amsfonts}

\usepackage{amssymb}

\usepackage[dvips]{graphicx}

\usepackage{latexsym}

\usepackage{subfigure}

\usepackage{dcolumn}

\usepackage{bm}

\begin{document}

\begin{small}

\baselineskip=0.8 cm

\title{{\bf Observational constraints on the dark energy and dark matter
mutual coupling}}

\author{Chang Feng}

\affiliation{Department of Physics, Fudan University, Shanghai
200433, China}

\author{Bin Wang}
\affiliation{Department of Physics, Fudan University, Shanghai
200433, China}

\author{Elcio Abdalla}
\affiliation{Instituto de Fisica, Universidade de Sao Paulo, CP
66318, 05315-970, Sao Paulo, Brazil}

\author{ Rukeng Su}

\affiliation{China Center of Advanced Science and Technology
(World Laboratory), P.B.Box 8730, Beijing 100080,
\\ Department of Physics, Fudan University, Shanghai 200433, China}

\vspace*{0.2cm}

\begin{abstract}

\baselineskip=0.8 cm

We examine different phenomenological interaction models for Dark
Energy and Dark Matter by performing statistical joint analysis
with observational data arising from the 182 Gold type Ia
supernova samples, the shift parameter of the Cosmic Microwave
Background given by the three-year Wilkinson Microwave Anisotropy
Probe observations, the baryon acoustic oscillation measurement
from the Sloan Digital Sky Survey and age estimates of 35
galaxies. Including the time-dependent observable, we add
sensitivity of measurement and give complementary results for the
fitting. The compatibility among three different data sets seem to
imply that the coupling between dark energy and dark matter is a
small positive value, which satisfies the requirement to solve the
coincidence problem and the second law of thermodynamics, being
compatible with previous estimates.

\end{abstract}

\pacs{98.80.Cq, 98.80-k } \maketitle

\newpage

\baselineskip=0.4 cm

Our universe is undergoing an accelerated expansion driven by a so
called dark energy (DE). Since DE occupies almost 70\% of the
energy content of the universe today, it is natural to consider
its interaction with the remaining fields of Standard model and
its generalizations. It has been claimed that the coupling between
DE and dark matter (DM) can provide a mechanism to alleviate the
coincidence problem \cite{2,3}. Furthermore, it has been argued
that an appropriate interaction between DE and DM can influence
the perturbation dynamics and affect the lowest multipoles of the
CMB spectrum \cite{4,5}. Recently, it has been shown that such an
interaction could be inferred from the expansion history of the
universe, as manifested in the supernova data together with CMB
and large-scale structure\cite{6}. However, the observational
limits on the strength of such an interaction remain weak\cite{7}.
Signature of the interaction between DE and DM in the dynamics of
galaxy clusters has also been analyzed \cite{8,9}. Other
discussions on the interaction between dark sectors can be found
in \cite{10,11,12}.

The interaction between DE and DM is a major issue to be
confronted in studying the physics of DE. However, since neither
DE nor DM is actually known to us, it is hard to describe the
interaction from first principles. Some attempts to discriminate
the interaction from the thermodynamical point of view have been
raised recently \cite{13a,13b}. Most studies on the interaction
between dark sectors rely either on the assumption of interacting
fields from the outset \cite{14,15}, or from phenomenological
requirements\cite{4,7,10,16}. In view of continuity equations, the
interaction between DE and DM must be a function of the energy
density multiplied by a quantity with units of inverse of time,
which can be chosen as the Hubble factor $H$. There is freedom to
choose the form of the energy density, which can be any
combination of DE and DM. Thus, the interaction between DE and DM
could be expressed phenomenologically in forms such as $Q =
Q(H\rho_{DM})$\cite{7,16}, $Q = Q(H\rho_{DE})$\cite{13b}, $Q =
Q(H(\rho_{DE} + \rho_{DM}))$\cite{4,10}.

It is of great interest to investigate effects of different forms
of interaction between DE and DM on the universe evolution. In
\cite{16} the impact of the interaction proportional to the DM
energy density on the determination of a redshift dependent DE
equation of state (EOS) and on the DM density today has been
studied from SNIa data. It has been shown that the presence of
such a coupling increases the tension between the CMB data from
the analysis of the shift parameter and SNIa data for realistic
values of the present DM density fraction. Recently, a statistical
joint analysis by using observational data coming from the new 182
Gold SNIa samples, the shift parameter of the CMB given by the
three-year WMAP observations and the baryonic acoustic oscillation
measurement from the SDSS has been carried out for the interaction
between DE and DM\cite{18}. Comparisons concerning the influence
on cosmological parameters and the effect on solving cosmic
coincidence problem among different forms of phenomenological
interaction models have been done. It was argued that consequences
of DE and DM interaction on cosmological parameters are sensitive
to the DE EOS. Choosing an appropriate EOS and the interaction in
proportional to the energy density of DE, a positive coupling
turns out to be more probable and the coincidence problem gets
alleviated. However, for other forms of the phenomenological
interaction models and for other DE EOS, one gets a negative
coupling between dark sectors which will result in unphysical
situations and fail to solve the coincidence problem. The negative
coupling has also been seen by using the same data from SNIa
together with CMB and large-scale structure for the interacting
holographic DE model with the interaction proportional to the
total dark sector energy density\cite{6} and other models
describing the interaction in proportional to the DM energy
density\cite{7}. It was argued that the negative coupling is not
able to alleviate the coincidence problem\cite{pavon} and the
model does not obey the second law of thermodynamics \cite{13b}.
Using the galaxy cluster data, the coupling was obtained to be
positive indicating the energy decay from DE to DM.

To reduce the uncertainty and put tighter constraint on the value
of the coupling between DE and DM, new observables should be
added. Recalling that the test of cosmological models by SNIa data
is a distance based method, it is of interest to look for tests
based on time-dependent observables. In \cite{5,19}, the age of an
old high redshift galaxy has been used to constrain the model.  In
this work we will combine four fundamental observables including
the new 182 Gold SNIa samples, the shift parameter of the CMB
given by the three-year WMAP observations, the baryon acoustic
oscillation (BAO) measurement from the SDSS and age estimates of
35 galaxies provided in\cite{alcaniz14} to perform the joint
systematic analysis of the coupling between dark sectors. We
expect that sensitivities of measurements of different observables
can give complementary results on the coupling between dark
sectors. We will compare the compatibility among SNIa data
including BAO, CMB and age data and determine the tendency of the
coupling results.

Concerning energy conservation for the overall energy density of
dark sectors, we can suppose that the interaction between DE and
DM is described by
\begin{equation}
\dot{\rho_m}+3H\rho_m=Q,
\end{equation}
\begin{equation}
\dot{\rho_D}+3H(1+\omega_D)\rho_D=-Q,
\end{equation}
where $Q$ denotes the interaction term.

From the equations above, phenomenological forms of the
interaction between DE and DM must be a function of the energy
densities multiplied by a quantity with units of inverse of time,
which have possible expressions, such as (1) $Q=\delta
H(\rho_{DM}+\rho_{DE})$, (2) $Q=\delta H\rho_{DM}$ and (3)
$Q=\delta H\rho_{DE}$ etc.

We will constrain the coupling between DE and DM in different
phenomenological interaction models by using the latest
observations (golden SN Ia, the shift parameter of CMB and the
BAO) and combining them with the lookback time data. We will not
specify any special model of DE. Considering recent accurate data
analysis showing that the time varying DE gives a better fit than
a cosmological constant and in particular, DE EoS can cross $-1$
around $z = 0.2$ from above to below\cite{21}, we will employ two
commonly used parameterizations in our work, namely
\begin{equation}
\omega_I(z)=w_0+\frac{w_1z}{(1+z)},
\end{equation}
\begin{equation}
\omega_{II}(z)=w_0+\frac{w_1z}{(1+z)^2}.
\end{equation}

The up-to-date gold SN Ia sample was compiled by Riess et al
\cite{feng10}. This sample consists of 182 data, which gives the
distance modulus at redshift $z$. The distance modulus is defined
as
\begin{equation}
\mu_{th}(z;\textbf{P},\tilde{M})=5\log_{10}(d_L(z)/{\rm Mpc})+25=
5\log_{10}[(1+z)\int_0^z\frac{dz^{\prime}}{E(z^{\prime})}]+25-5\log_{10}H_0,
\end{equation}
where the luminosity distance
$d_L(z)=\frac{c(1+z)}{H_0}\int_0^z\frac{dz^{\prime}}{E(z^{\prime})}$,
the nuisance parameter $\tilde{M}=5\log_{10}H_0$ is marginalized
over by assuming a flat prior $P(H_0)=1$ on $H_0$ and $\textbf{P}$
describes a set of parameters characterizing the given model.

An efficient way to reduce the degeneracies of the cosmological
parameters is to use the SN Ia data in combination with the BAO
measurement from SDSS \cite{feng12} and the CMB shift parameter
\cite{feng11}.  Using a large sample of 46748 luminous red
galaxies covering 3816 square degrees out to a redshift of
$z=0.47$ from the SDSS, Eisenstein et al \cite{feng12} have found
the model independent BAO measurement which is described by the
$A$ parameter
\begin{equation}
A=\sqrt{\Omega_m}E(z_{BAO})^{-1/3}[\frac{1}{z_{BAO}}
\int_0^{z_{BAO}}\frac{dz^{\prime}}{E(z^{\prime})}]^{2/3}\\
=0.469(\frac{n_s}{0.98})^{-0.35}\pm0.017,
\end{equation}
where $n_s$ can be taken as $0.95$ \cite{fengWMAP3y} and
$z_{BAO}=0.35$.

The CMB shift parameter is given by
\begin{equation}
R=\sqrt{\Omega_m}\int_0^{z_{ls}}\frac{dz^{\prime}}{E(z^{\prime})},
\end{equation}
where $z_{ls}=1089$. This CMB shift parameter $R$ captures how the
$l$-space positions of the acoustic peaks in the angular power
spectrum shift. Its value is expected to be the least model
independent and can be extracted from the CMB data. The WMAP3 data
\cite{fengWMAP3y} gives $R=1.70\pm 0.03$ \cite{feng11}.

We now turn to the lookback time observations. They have been used
in \cite{zhu} and have been shown effective to provide a
complementary test of different models. By assuming the total age
of the universe to be $t_0 = 13.7\pm 0.2 Gyr$, as given by current
CMB measurement \cite{fengWMAP3y}, we transform the age estimates
of 35 galaxies provided in \cite{alcaniz14}. The lookback
time-redshift relation is defined by
\begin{equation}
t_L(z;\textbf{P})=H_0^{-1}\int_0^z\frac{dz^{\prime}}{(1+z^{\prime})E(z^{\prime})},
\end{equation}
where $H_0^{-1}=9.78h^{-1}Gyr$. We have adopted the recent value
0.72 for $h$ given by the HST key project \cite{feng23},
$\textbf{P}$ stands for the model parameters. To use the lookback
time and the age of the universe to test a given cosmological
model, we follow \cite{feng8} and consider an object $i$ whose age
$t_i(z)$ at redshift $z$ is the difference between the age of the
universe when it was born at redshift $z_F$ and the universe age
at $z$,
\begin{equation}
t_i(z)=H_0^{-1}[\int_{z_i}^{\infty}\frac{dz^{\prime}}{(1+z^{\prime})E(z^{\prime})}
-\int_{z_F}^{\infty}\frac{dz^{\prime}}{(1+z^{\prime})E(z^{\prime})}].
\end{equation}
Using the lookback time definition, we have
$t(z_i)=t_L(z_F)-t_L(z)$. Thus the lookback time to an object at
$z_i$ can be expressed as
\begin{equation}
t_L^{obs}(z_i)=t_L(z_F)-t(z_i)=[t_o^{obs}-t_i(z)]-[t_o^{obs}-t_L(z_F)]=t_o^{obs}-t_i(z)-df,
\end{equation}
where $df=t_o^{obs}-t_L(z_F)$ is the delay factor.

We employ the Monte-Carlo Markov Chain (MCMC) method \cite{he36}
to explore the parameter space. By using two parameterizations for
the EoS of DE $\omega_I , \omega_{II}$, we show in table 1 the
parameter space when the coupling between DE and DM is taken
proportional to energy densities of DM, DE, total DM plus DE (T),
respectively.

Comparing with the result obtained in \cite{18}, it is interesting
to find that by adding the new observable, the lookback time data,
it is possible to have positive coupling between DE and DM,
especially for the EOS with the form $\omega_{II}$.

\tabcolsep0.2in

Table 1.

\begin{small}PARAMETERS AT $68.3\%$ CONFIDENCE LEVEL\\
\begin{tabular}{cccccc}
\hline \hline
Coupling&EoS & $w_0$ & $w_1$ & $\Omega_m$ & $\delta$\\
\hline T&$\omega_I$ &$-1.10^{+0.15}_{-0.15}$&$1.22^{+
0.20}_{-0.27}$&$0.27^{+0.02}_{-0.02}$&$-0.01^{+0.07}_{-0.04}$\\
T&$\omega_{II}$&$-1.50^{+0.31}_{-0.30}$&$3.90^{+2.09}_{-2.31}$&$0.26^{+0.02}_{-0.02}$&
$0.01^{+0.03}_{-0.03}$\\
\hline

DM&$\omega_I$ &$-1.17^{+
0.16}_{-0.14}$&$1.28^{+0.21}_{-0.32}$&$0.27^{+0.02}_{-0.02}$&$-0.02^{+0.31}_{-0.06}$\\
DM&$\omega_{II}$&$-1.50^{+0.32}_{-0.31}$&$3.91^{+2.12}_{-2.34}$&$0.25^{+0.02}_{-0.02}$&$0.01^{+0.04}_{-0.04}$\\
\hline

DE&$\omega_I$
&$-1.11^{+0.17}_{-0.16}$&$1.19^{+0.19}_{-0.28}$&$0.27^{+0.01}_{-0.01}$&$-0.04^{+0.13}_{-0.13}$\\
DE&$\omega_{II}$&$-1.49^{+0.31}_{-0.30}$&$3.78^{+2.13}_{-2.39}$&$0.26^{+0.02}_{-0.02}$&$0.05^{+0.06}_{-0.10}$\\
\hline
\end{tabular} \\
\end{small}
\begin{scriptsize}Note --- (1)The flat priors on the parameters for EoS1
are, $-10<w_0<10$, $-10<w_1<10$, $0<\Omega_m<0.8$,
$-0.5<\delta<0.5$;(2)For EoS2 ,$-10<w_0<10$,$-15<w_1<15$,
$0<\Omega_m<0.8$, $-1<\delta<1$. The CMBEASY GUI is utilized to
process the MCMC chains.
\end{scriptsize}\\

  \tabcolsep0.2in
Table 2.\\
\begin{small}PARAMETERS AT $68.3\%$ CONFIDENCE LEVEL\\
\begin{tabular}{ccccc}
\hline \hline
Coupling&EoS & $w_0$ & $\Omega_m$ & $\delta$\\
\hline T&$\omega_I$
&$-1.13^{+0.02}_{-0.08}$&$0.26^{+0.01}_{-0.01}$&$0.02^{+0.18}_{-0.03}$\\
\hline
T&$\omega_{II}$&$-1.50^{+0.07}_{-0.08}$&$0.25^{+0.02}_{-0.01}$&$0.01^{+0.02}_{-0.02}$\\

\hline
DM&$\omega_I$&$-1.22^{+0.04}_{-0.06}$&$0.26^{+0.01}_{-0.01}$&$0.04^{+0.16}_{-0.02}$\\
\hline
DM&$\omega_{II}$&$-1.50^{+0.07}_{-0.08}$&$0.25^{+0.02}_{-0.02}$&$0.01^{+0.03}_{-0.03}$\\

\hline
DE&$\omega_I$&$-1.18^{+0.04}_{-0.12}$&$0.26^{+0.01}_{-0.01}$&$0.06^{+0.14}_{-0.09}$\\
\hline
DE&$\omega_{II}$&$-1.48^{+0.08}_{-0.10}$&$0.26^{+0.01}_{-0.01}$&$0.05^{+0.06}_{-0.08}$\\

\hline
\end{tabular}\\
\end{small}
\begin{scriptsize}Note --- The flat priors are, $-5<w_0<5$,
$0<\Omega_m<0.5$, $-0.2<\delta<0.2$.
\end{scriptsize}

\begin{figure}
\renewcommand{\captionfont}{\scriptsize}
\centering
\includegraphics[width=7cm,height=6cm]{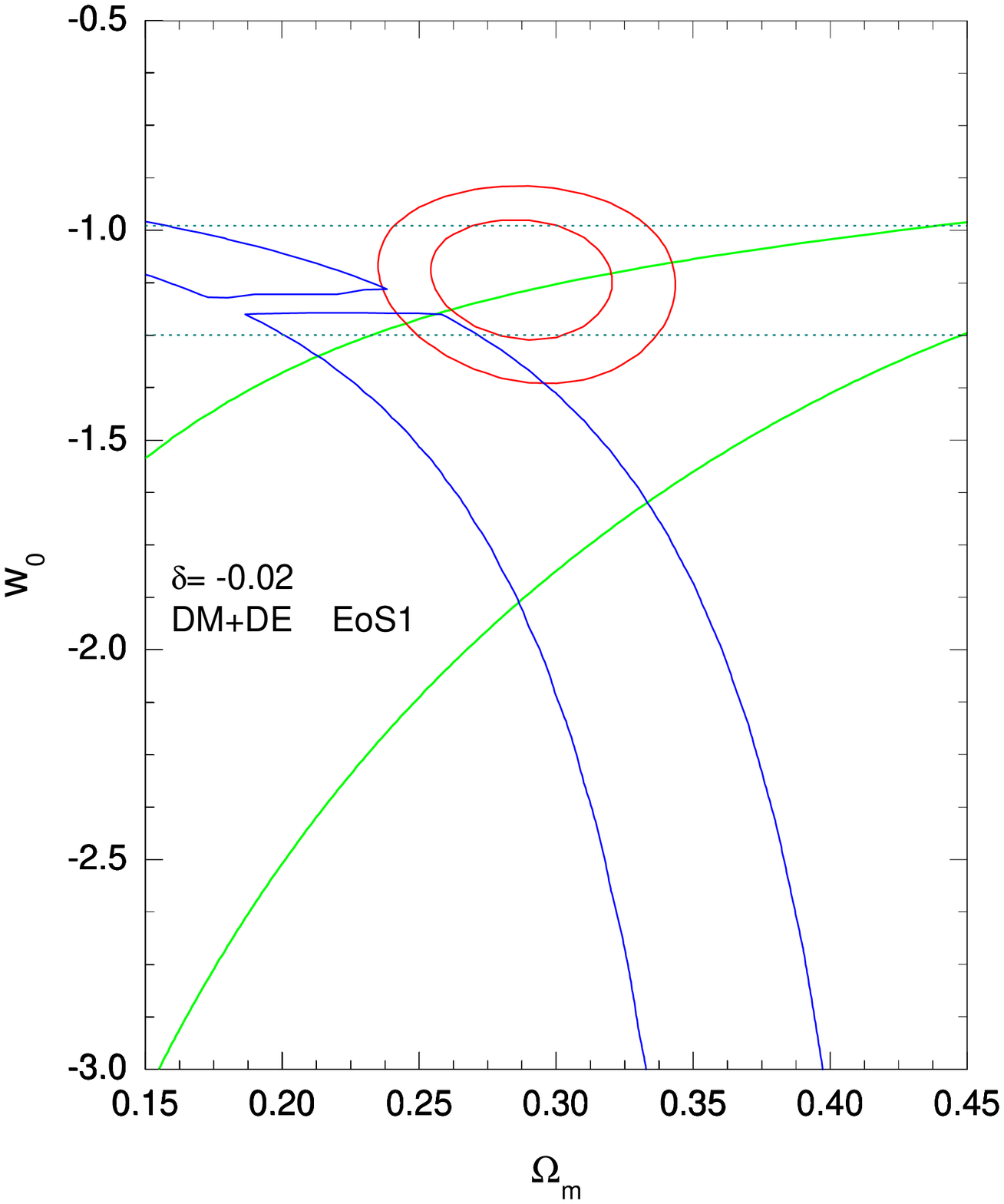}
\includegraphics[width=7cm,height=6cm]{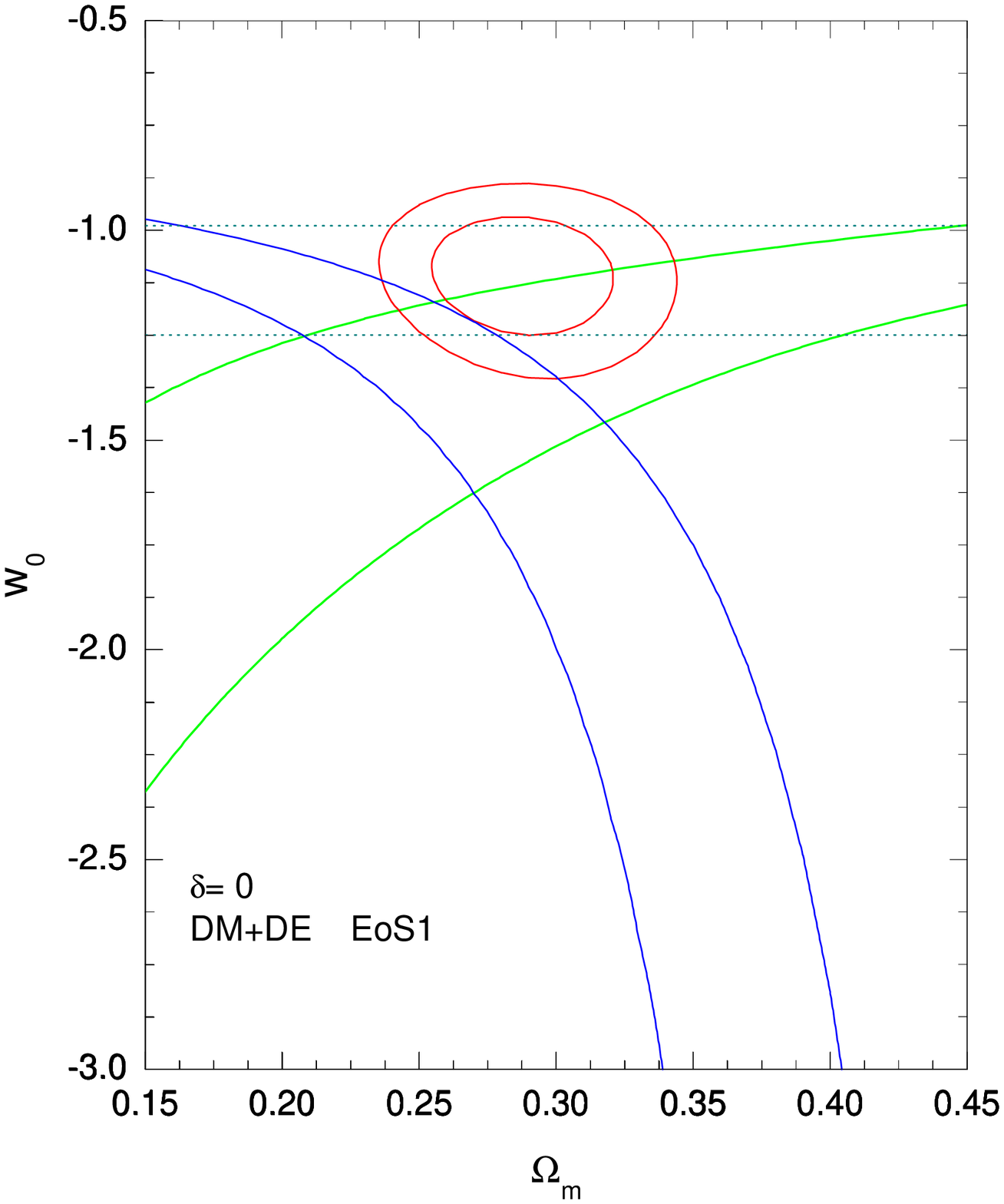}
\includegraphics[width=7cm,height=6cm]{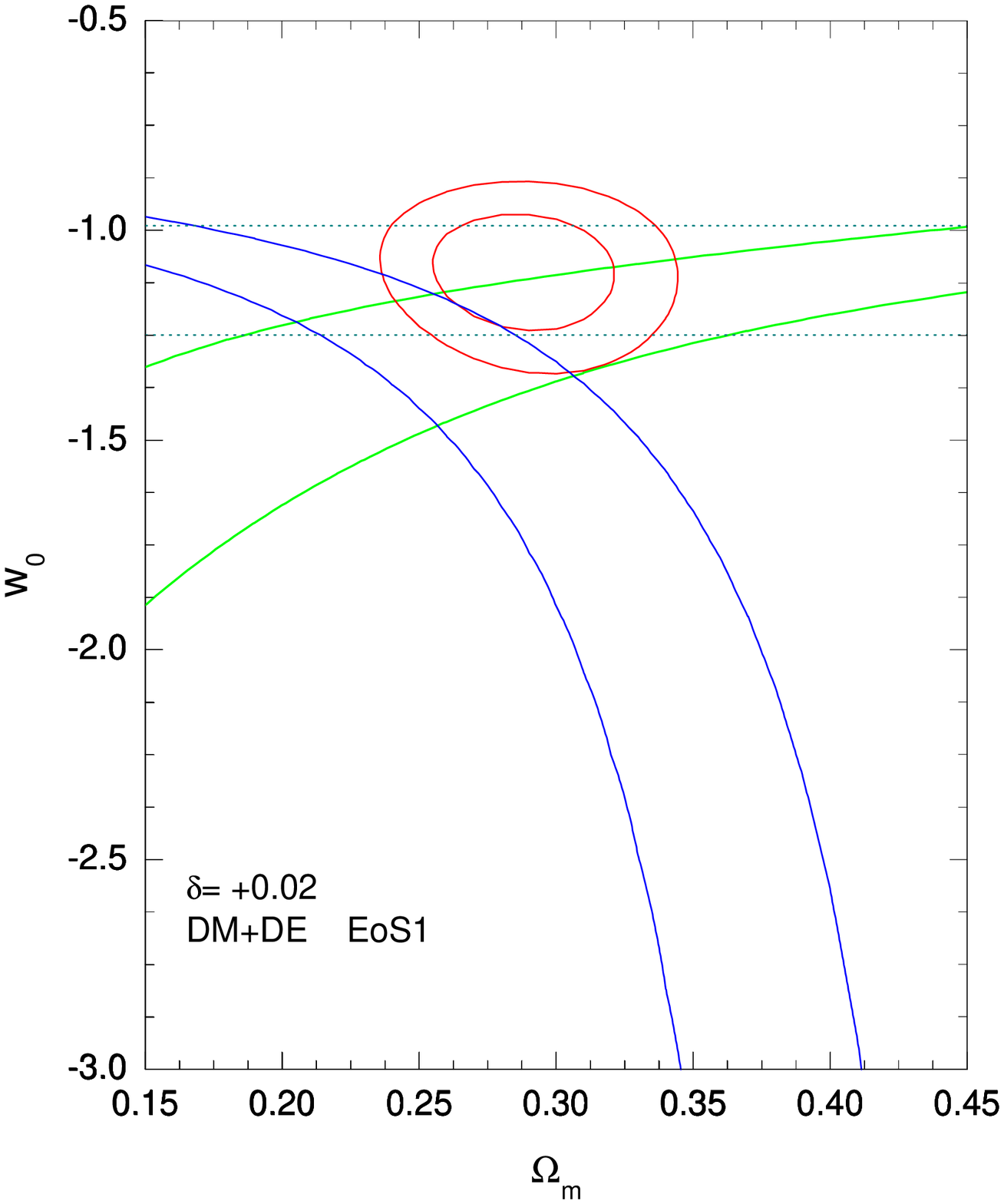}
\includegraphics[width=7cm,height=6cm]{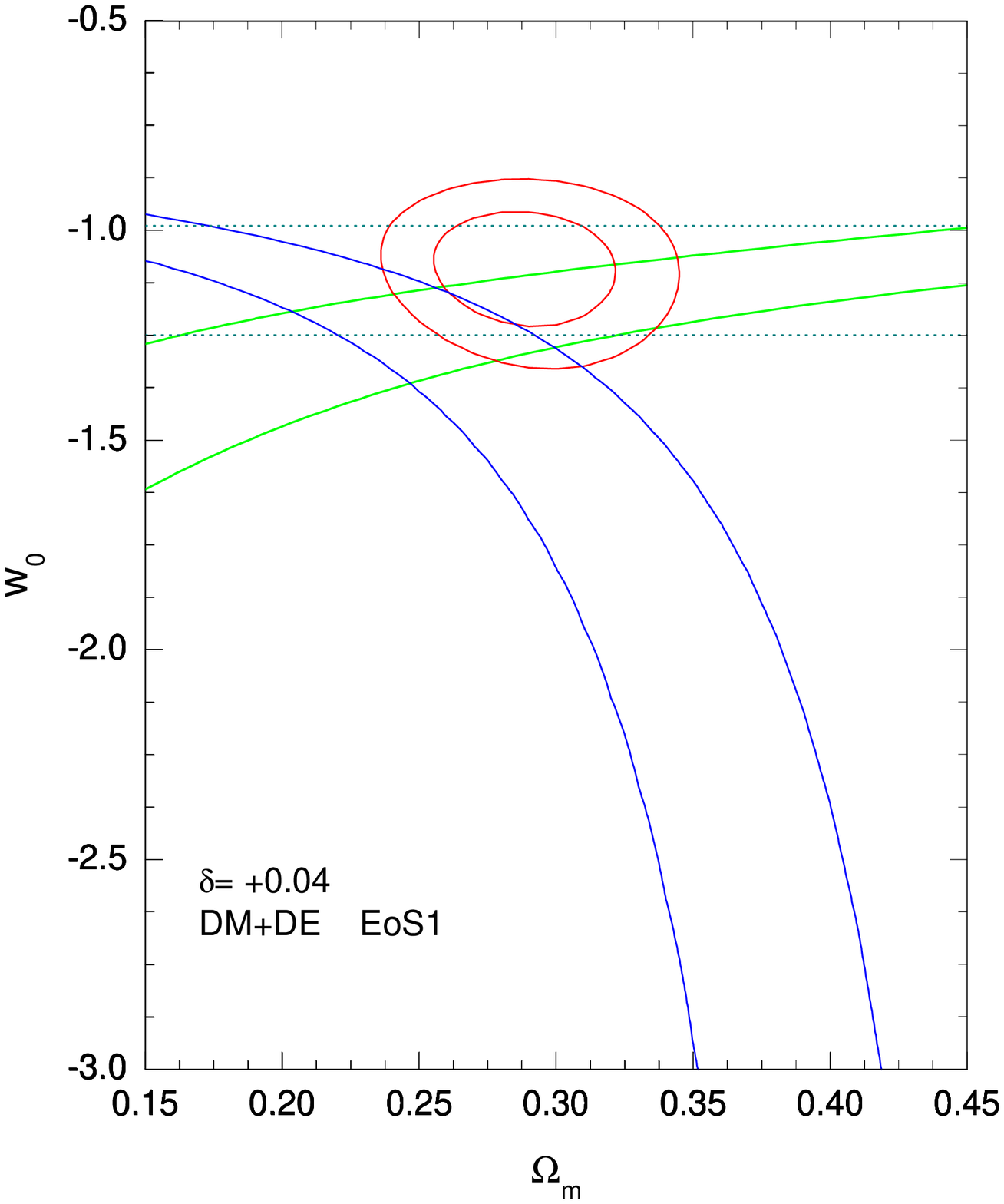}
 \caption{Contour plots in the $\Omega_m$-$w_0$ plane for a
variable EoS $\omega_I$ when Q is proportional to the total energy
density of DM and DE after giving a prior to $w_1=1.22$. The
compatibility among CMB , SNIa+BAO, Lookback time can be compared by
examining the 2$\sigma$ contour of CMB shift constraint(green line),
the Lookback time constraint(blue line), and 1$\sigma$, 2$\sigma$
contours of SNIa+BAO result(red line). The 5 years WMAP results for
$w_0$ are also indicated by parallel lines.}
\end{figure}

\begin{figure}
\renewcommand{\captionfont}{\scriptsize}
\centering
\includegraphics[width=7cm,height=6cm]{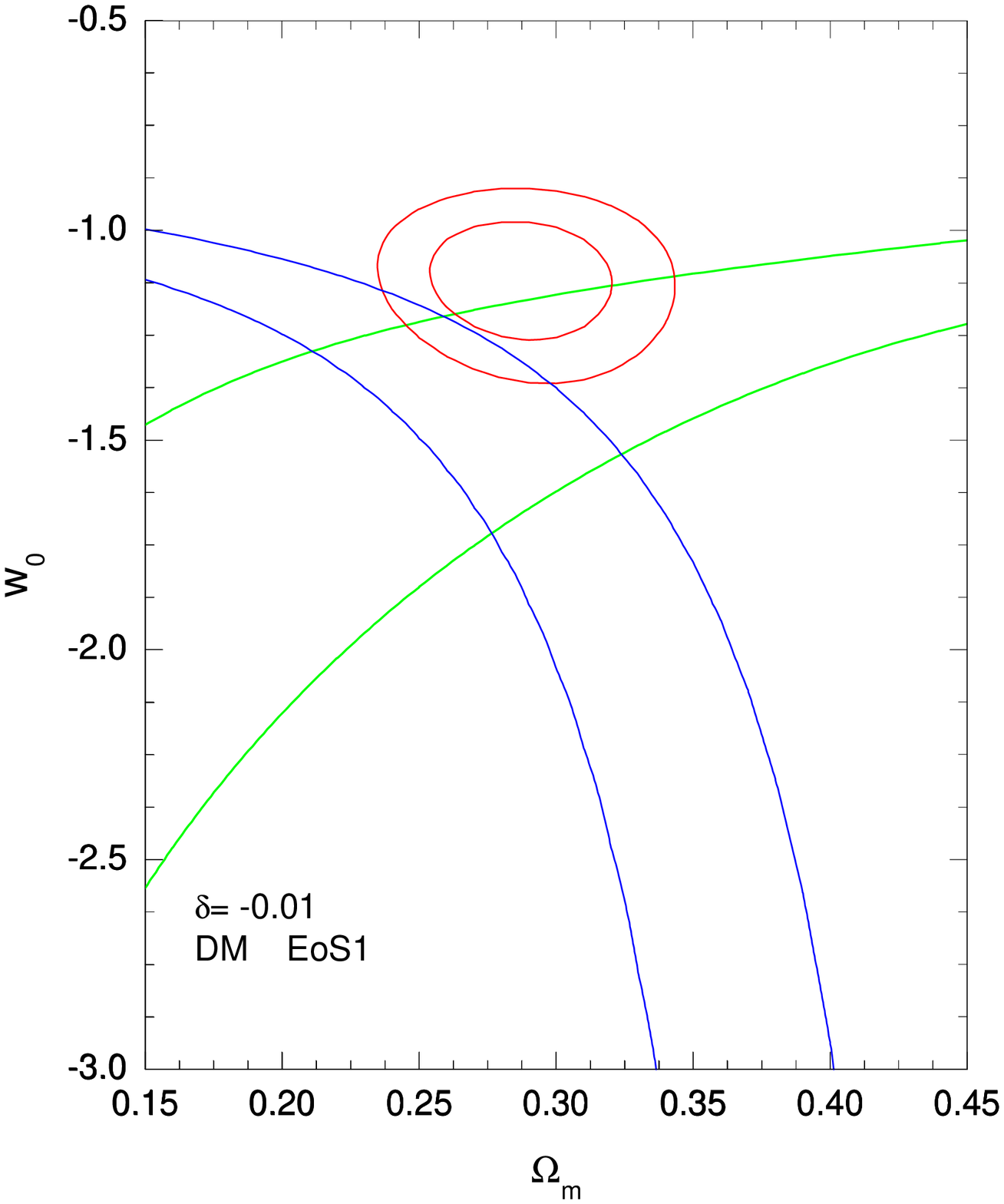}
\includegraphics[width=7cm,height=6cm]{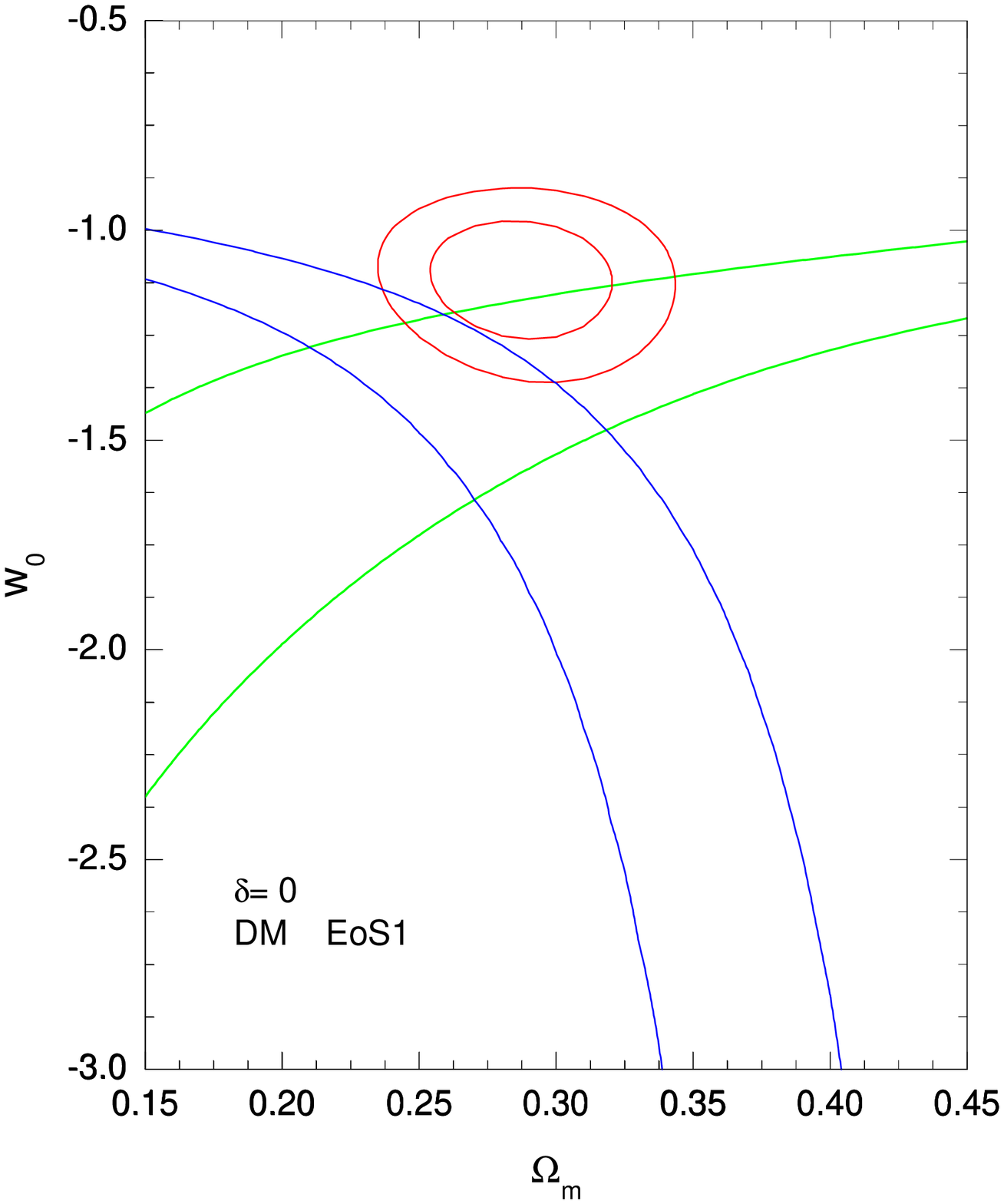}
\includegraphics[width=7cm,height=6cm]{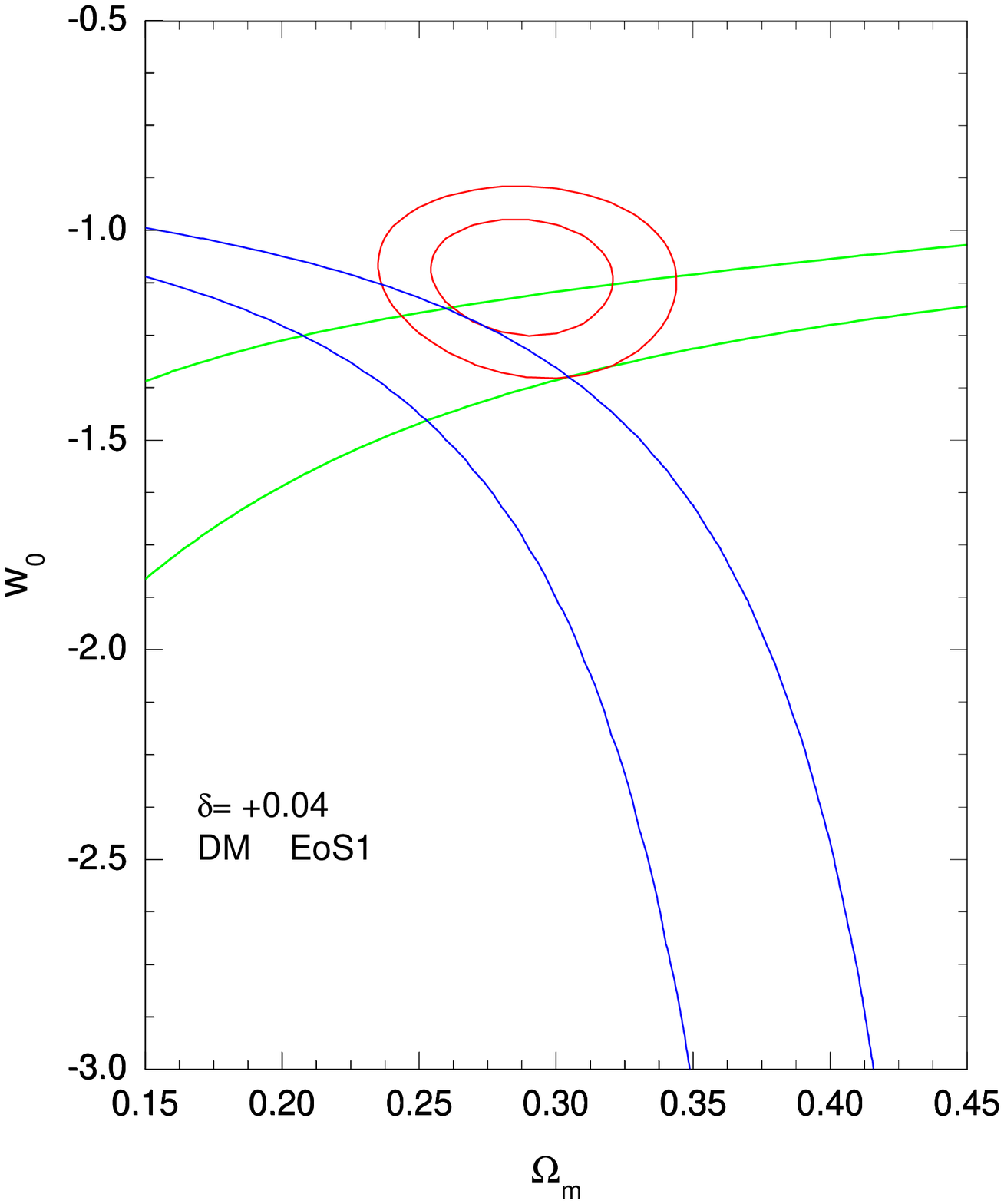}
\includegraphics[width=7cm,height=6cm]{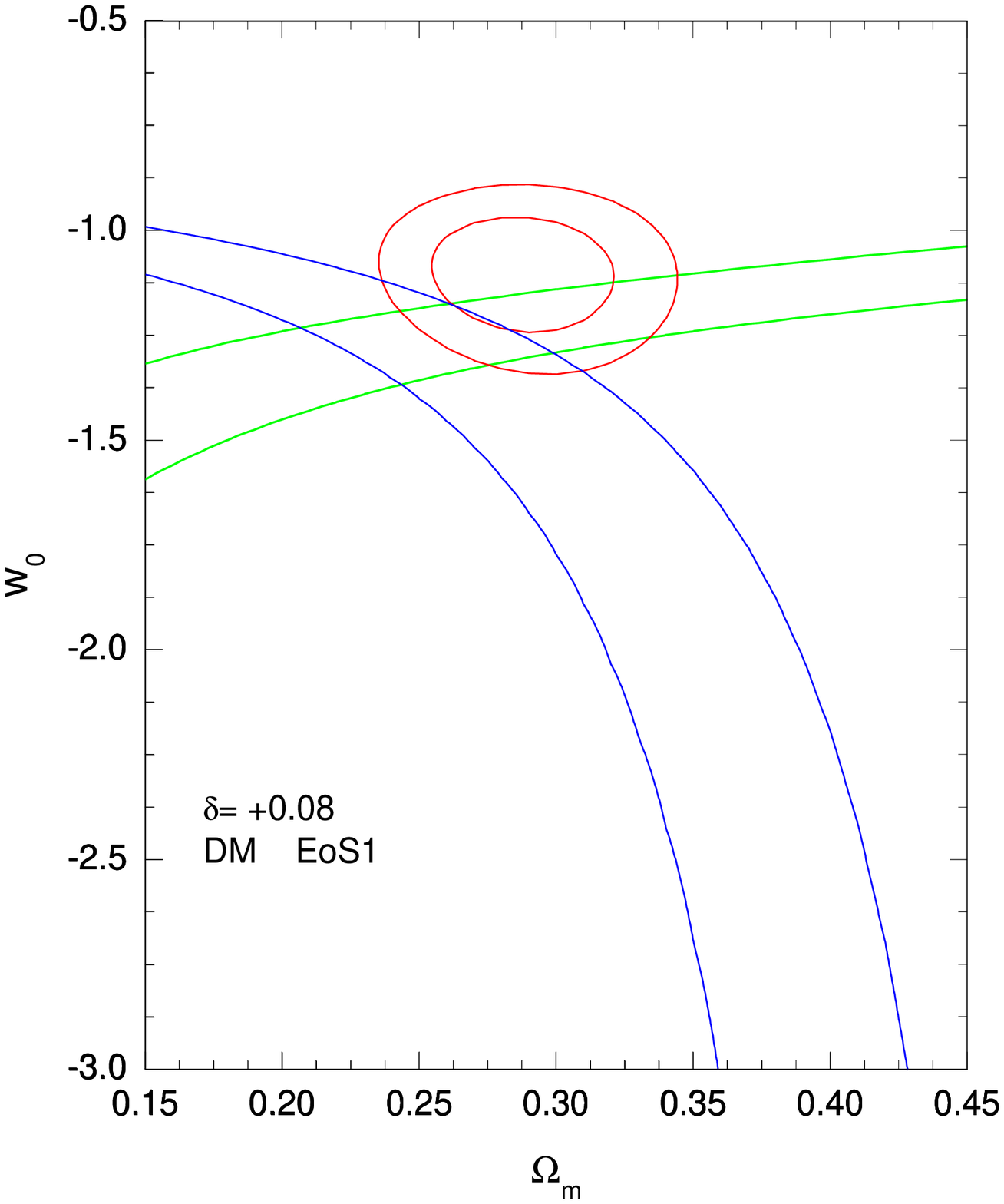}
\caption{Contour plots in the $\Omega_m$-$w_0$ plane for a
variable EoS $\omega_I$ when Q is proportional the energy density
of DM after giving a prior to $w_1=1.28$. Green lines are for
2$\sigma$ contours of CMB shift constraint, blue lines are for the
Lookback time constraints and red lines are for 1$\sigma$
,2$\sigma$ contours of SNIa+BAO results.}
\end{figure}

\begin{figure}
\renewcommand{\captionfont}{\scriptsize}
\centering
\includegraphics[width=7cm,height=6cm]{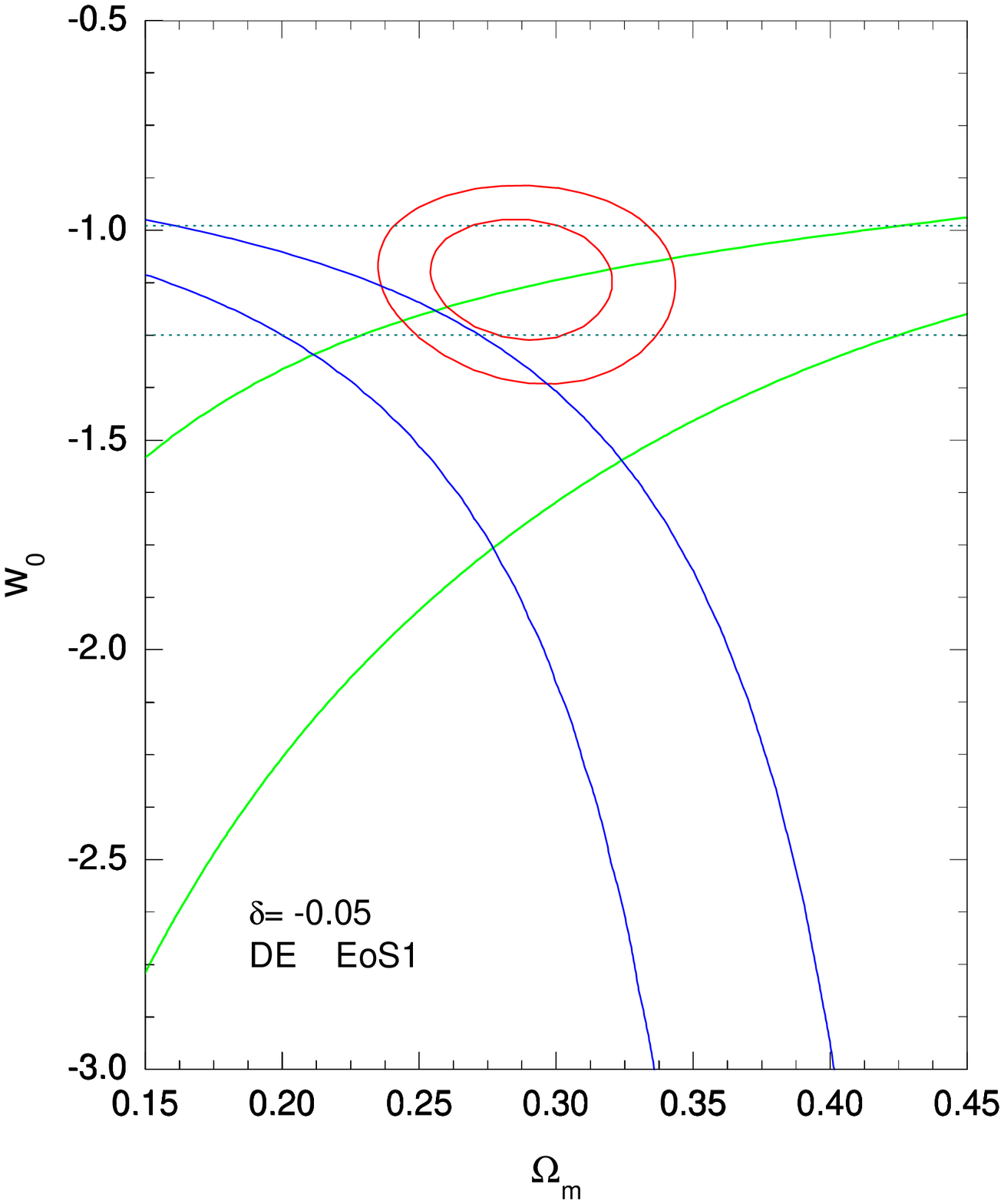}
\includegraphics[width=7cm,height=6cm]{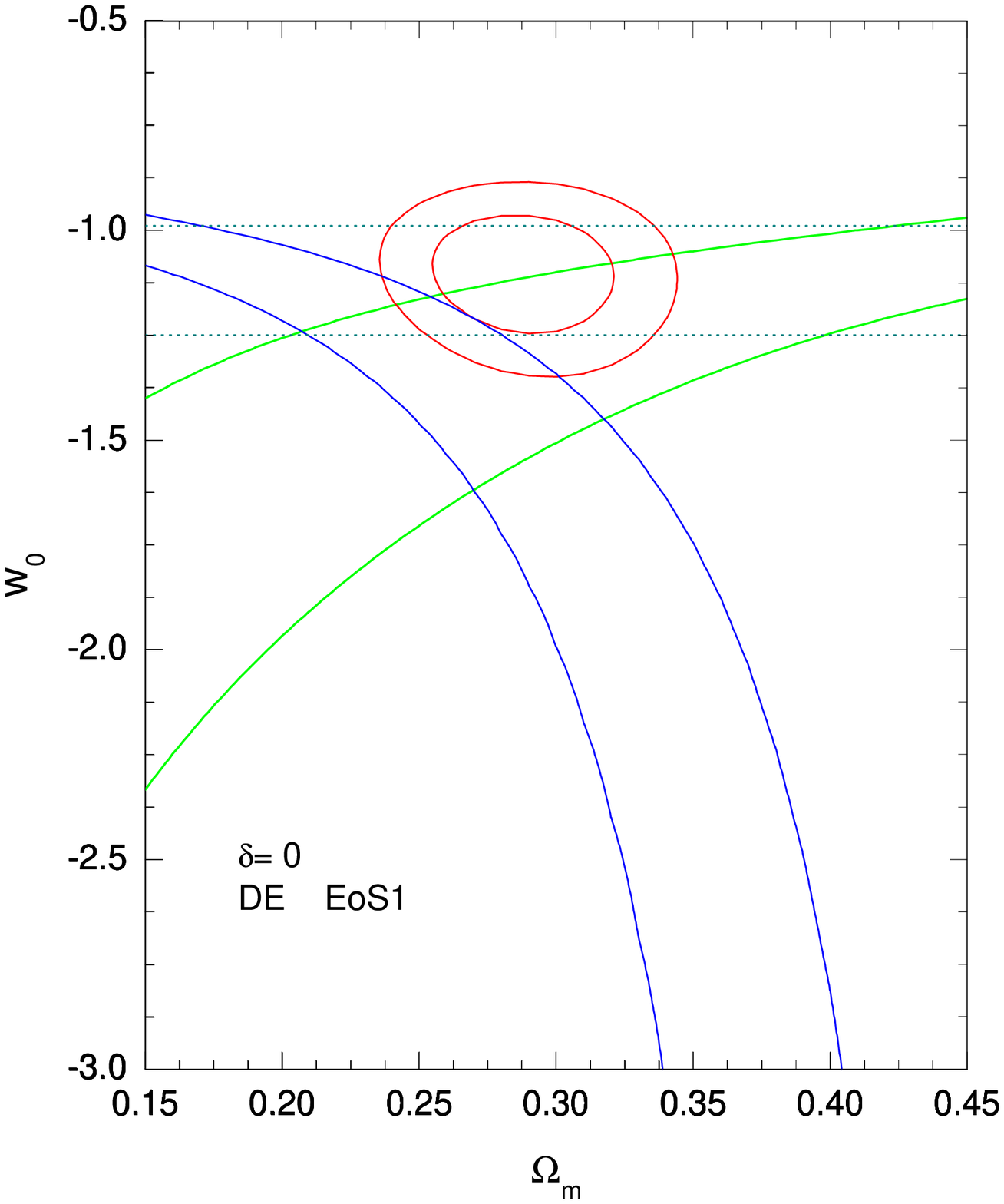}
\includegraphics[width=7cm,height=6cm]{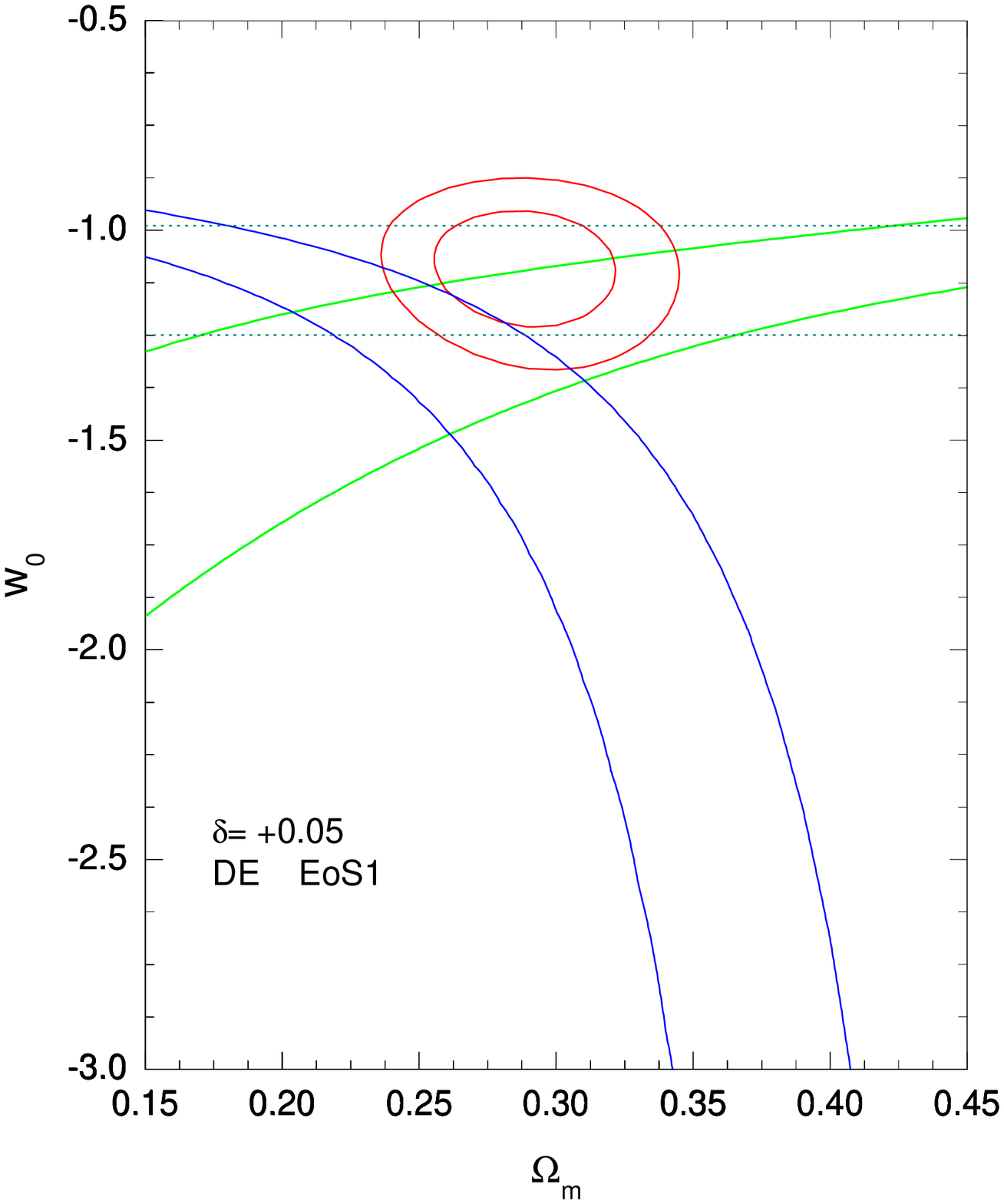}
\includegraphics[width=7cm,height=6cm]{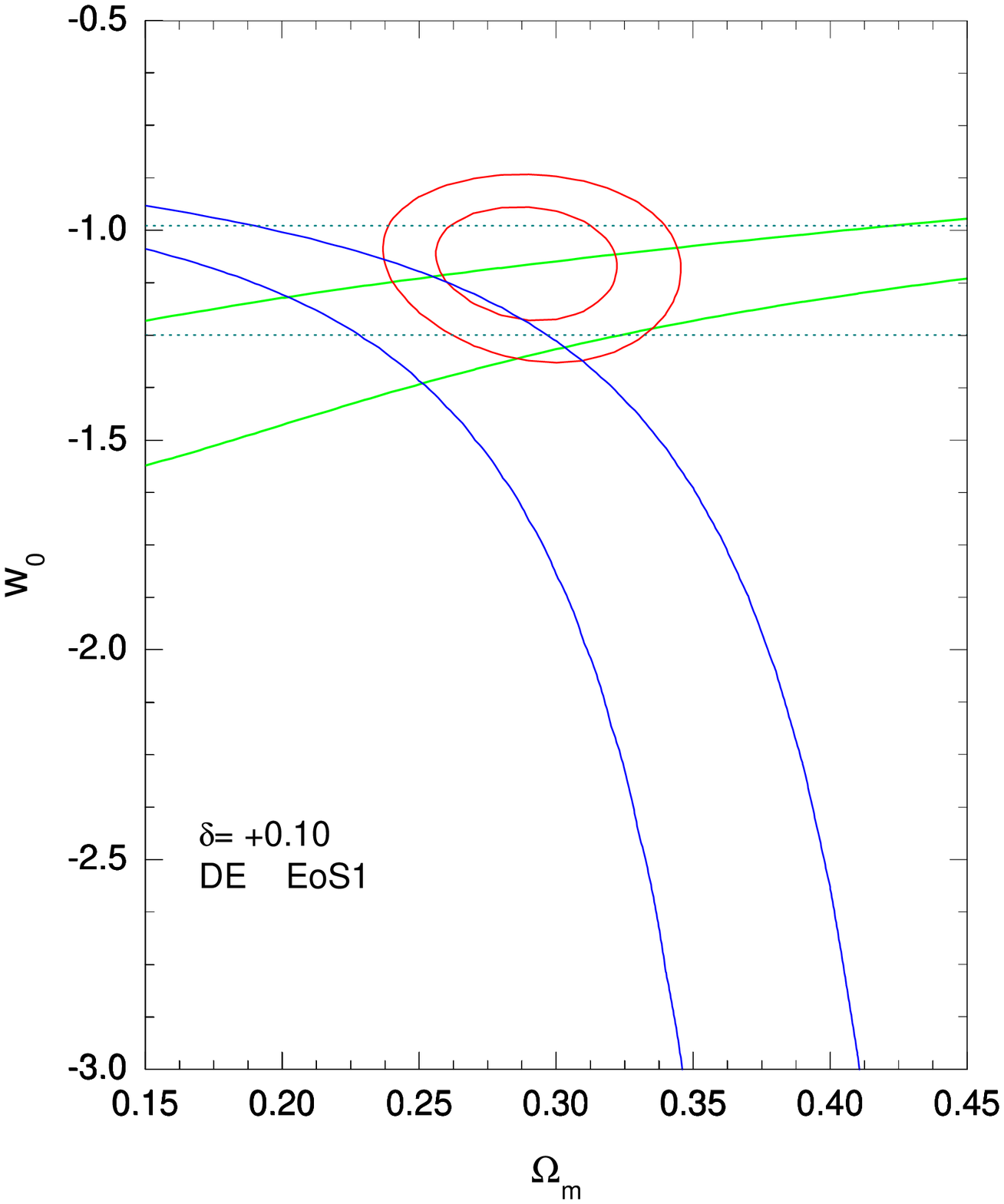}
\caption{Contour plots in the $\Omega_m$-$w_0$ plane for a
variable EoS $\omega_I$ when Q is in proportional to the DE energy
density after giving a prior to $w_1=1.19$. Green lines are for
2$\sigma$ contours of CMB shift constraint, blue lines are for the
Lookback time constraints and red lines are for 1$\sigma$
,2$\sigma$ contours of SNIa+BAO results. The 5 years WMAP results
for $w_0$ are also indicated by parallel lines. }
\end{figure}

We perform the data comparisons for different phenomenological
interaction models between DE and DM with two different
parameterizations of DE EOS. By taking priority of $w_1$ obtained
from MCMC fitting as central values, we plot the contours in
$\Omega_{DM}-w_0$ plane for different interaction models in Figs
1, 2, 3  with DE EOS $\omega_I$. Similar results for DE EOS
$\omega_{II}$ have also been observed. Green lines indicate the
result from CMB shift parameters and blue lines are from lookback
time result.

We observe that when $|\delta|$ is over the range in table 1,
there appear poorer compatibility among the three data sets,
especially between CMB and SNIa data. For the big positive
$\delta$, this incompatibility for the interaction between dark
sectors proportional to the DM energy density was also observed in
\cite{16}. For small enough $|\delta|$, the range obtained from
CMB shift parameters will change more compared to constraints from
other two data sets. The lower green line for the CMB shift moves
upper when small $\delta$ becomes more positive and the upper
green line becomes more flattened. This leads more overlapped
region for three data sets in the $\Omega_{DM}-w_0$ plane for
small positive value of the coupling. However when the positive
coupling is over a limit, the lower green line will cut the
contour, while the upper green line cannot efficiently move upper,
which will reduce the overlapped region of the constraints from
three different data sets. This result holds for all forms of
phenomenological interaction models and different
parameterizations of DE EOS.

Besides, in the small $|\delta|$ range, when $\delta$ becomes more
positive, we observed that there are more overlaps between
constraints from the SNIa+BAO and lookback time data sets. Thus
from the compatibility of three different data sets we obtain the
tendency of small positive coupling between DE and DM.

Choosing now the priority of $w_1$ as the central value from MCMC,
we obtain the parameter space listed in table 2. Using the
best-fit results of these parameters, we study the coincidence
problem. We pay attention to the ratio of energy densities between
DE and DM, $r=\rho_{DM}/\rho_{DE}$, and its evolution. In Fig.4 we
show the behavior of $r$ for the interaction between dark sectors
in proportional to DM energy density when we choose DE EOS to be
$\omega_{II}$. For other interaction forms and for DE EOS in the
form of $\omega_{I}, \omega_{II}$, $r$ behaviors are similar. We
see that with the positive coupling obtained from the best-fit
leads to a slower change of $r$ as compared to the noninteracting
case. This means that the period when energy densities of DE and
DM are comparable is longer compared to the noninteracting case.
Thus it is not so strange that we now live in the coincidence
state of the universe. In this sense the coincidence problem is
less acute when compared with the case without interaction.
Similar argument has also been given in \cite{li10}.

It is also worthwhile commenting on the results in the light of
the recent 5 years WMAP results \cite{5yearswmap}. We included, in
figures 1 and 3 the limits for $w_0$ from 5 years WMAP results.
They turn out to be perfectly compatible with the SNIa+BAO
countours at 1$\sigma$. This implies confidence in the results of
the present paper. For positive coupling, it has more possibility
for the overlapped region among three data sets to accommodate
$w_0$ within the 5 years WMAP region, which gives further strength
to the claims concerning the sign of the interaction.

\begin{figure}
\renewcommand{\captionfont}{\scriptsize}
\centering
\includegraphics[width=8cm,height=6cm]{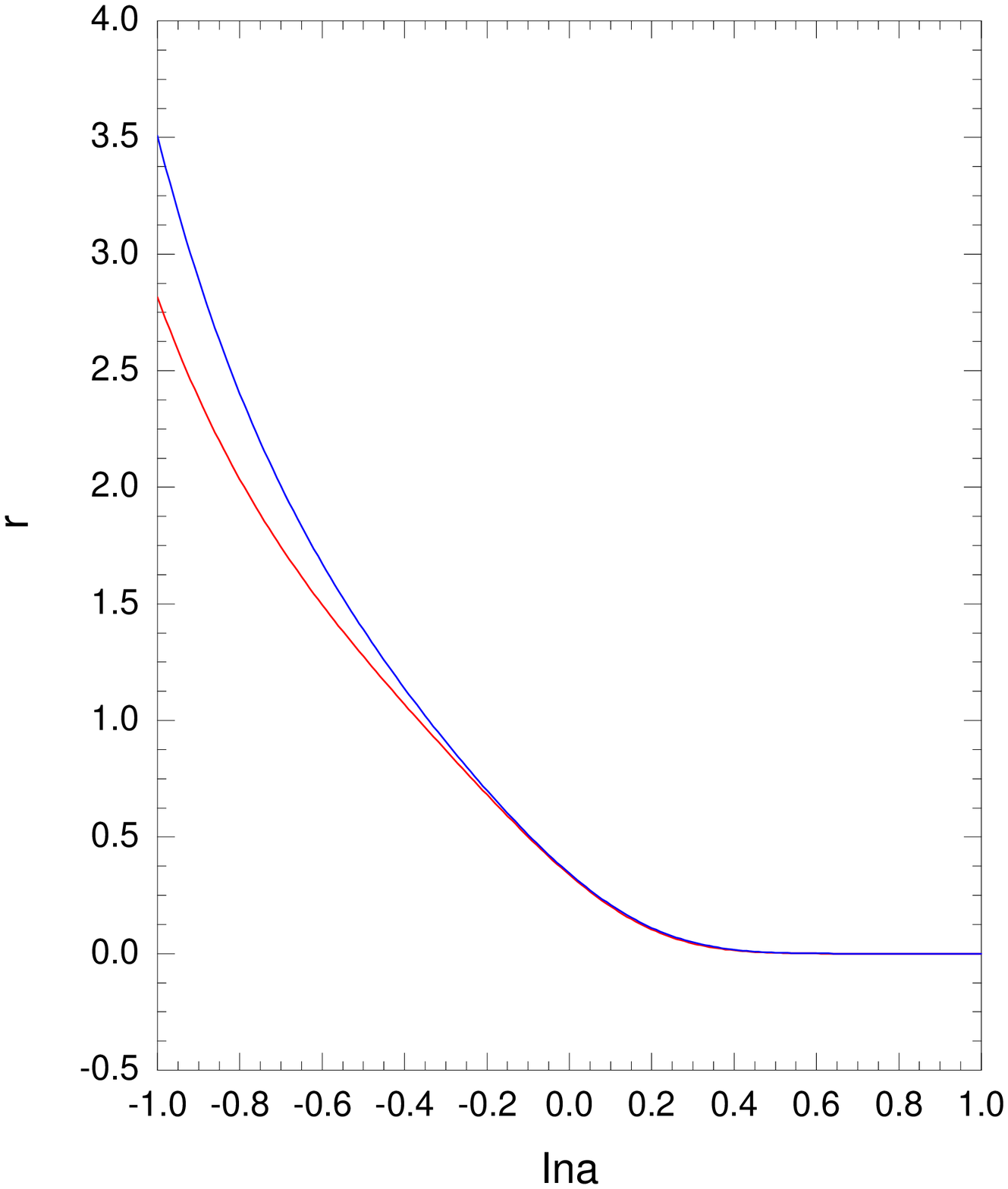}
\caption{The red line indicates the evolution of the ratio of
energy densities between DE and DM when the interaction is in
proportional to the energy density of DM and DE EoS is in the form
of $\omega_{II}$. We have compared the interacting case(red line)
with the non-interacting case(blue line).}
\end{figure}

In summary, we have examined different phenomenological
interaction models between DE and DM by performing statistical
joint analysis with observational data from the new 182 Gold SNIa
samples, the shift parameter of the CMB given by the three-year
WMAP, the baryon acoustic oscillation measurement from the SDSS
and age estimates of 35 galaxies. Comparing with the test by just
using data from SNIa together with CMB and large-scale
structure\cite{18}, we observed that adding the age constraint, we
get a tendency towards a positive coupling between DE and DM,
especially for the DE EOS with the form $\omega_{II}$. This shows
that the new observable can add sensitivity of measurement and
give a complementary result for the fitting.

We have studied the compatibility among three different data sets
including SNIa plus BAO, CMB shift and lookback time. We found
that the bigger couplings $|\delta|$ between dark sectors lead to
a poorer compatibility, especially comparing CMB with other two
data sets. For small $|\delta|$, we observed, for all
phenomenological forms of interaction with two parameterizations
of DE EOS,the same tendencies for the $\delta$ to be a small
positive number. The small positive coupling result is consistent
with that got independently by galaxy cluster analysis \cite{9}.
The positive coupling is required to alleviate the coincidence
problem and avoid some unphysical problems met in \cite{16,18}. It
is also the requirement of the second law of thermodynamics
\cite{13b}.

\begin{acknowledgments}

This work was partially supported by the NNSF of China, Shanghai
Education Commission, Science and Technology Commission; FAPESP
and CNPQ of Brazil.
\end{acknowledgments}

\end{small}

\end{document}